\documentclass[letterpaper]{article} % DO NOT CHANGE THIS
\usepackage[preprint]{aaai2027}  % Preprint mode for public arXiv distribution
\usepackage[hyphens]{url}  % DO NOT CHANGE THIS
\usepackage{graphicx} % DO NOT CHANGE THIS
\usepackage{natbib}  % DO NOT CHANGE THIS AND DO NOT ADD ANY OPTIONS TO IT
\usepackage{caption} % DO NOT CHANGE THIS AND DO NOT ADD ANY OPTIONS TO IT
\usepackage{amsmath,amssymb} % Required for displayed equations and math symbols
\usepackage{algorithm}
\usepackage{algorithmic}

\usepackage{newfloat}
\usepackage{listings}
\DeclareCaptionStyle{ruled}{labelfont=normalfont,labelsep=colon,strut=off} % DO NOT CHANGE THIS
\floatstyle{ruled}
\newfloat{listing}{tb}{lst}{}
\floatname{listing}{Listing}

\usepackage{booktabs}

\definecolor{bestblue}{RGB}{18,82,148}
\newcommand{\toolname}{\textsc{GraphAlignCoder}}
\newcommand{\best}[1]{\textbf{\textcolor{bestblue}{#1}}}
\newcommand{\score}[2]{#1 (#2)}

\title{GraphAlignCoder: Aligning Program and Proof Graphs for Code Generation}
\author{
    Yueke Zhang,
    Zihan Fang,
    Kevin Leach,
    Yu Huang
}
\affiliations{
    Vanderbilt University\\
    Nashville, TN 37015 USA\\
}

\begin{document}

\maketitle

\begin{abstract}
Code large language models (LLMs) can generate syntactically plausible programs that nevertheless violate hidden semantic constraints. Existing execution-feedback training methods identify whether a completed program fails, but provide limited supervision about how a correct solution should be organized. 
We introduce \toolname{}, a training framework that transfers explicit correctness structure into code generation. 

%\toolname{} constructs an implementation graph that captures control and dependence among program regions. In parallel, a constrained Lean pipeline produces kernel-checked proof traces, from which we extract a formal proof-flow graph.  The model first learns executable code together with graph-derived descriptions of why individual program regions are correct, then returns to Python-only generation. Under this single-sample greedy protocol, \toolname{} solves 50/175 LiveCodeBench v6 tasks, 23/148 BigCodeBench Hard tasks, and 363/1140 BigCodeBench Full tasks, outperforming both code-only SFT and CodeRL. The ablation study further shows that structural injection produces the initial reasoning gain, while consolidation and execution grounding are essential for robust cross-benchmark transfer. 

\toolname{} constructs an implementation graph that captures control and dependence among program regions. In parallel, a constrained Lean pipeline produces proof traces, from which we extract a formal proof-flow graph. The model first learns executable code together with graph-derived descriptions of why individual program regions are correct, and then consolidates this knowledge into code generation. \toolname{} consistently outperforms the base model, code-only SFT, and CodeRL across all benchmarks. Compared with CodeRL, it increases the solved count from 38 to 50 on LiveCodeBench v6 and from 16 to 23 on BigCodeBench Hard, corresponding to relative gains of 31.6\% and 43.8\%, while also improving BigCodeBench Full from 359 to 363 tasks. The ablation study further shows that verification-graph injection produces the initial reasoning gain, while verification to code consolidation is essential for robust cross-benchmark transfer. 
\end{abstract}

\section{Introduction}
Large language models (LLMs) for code have rapidly moved from research prototypes to widely used programming assistants~\cite{chen2021codex,li2022alphacode,roziere2023codellama,lozhkov2024starcoder2,guo2024deepseekcoder,hui2024qwen25coder}.
These advances have also changed how programming is practiced: developers increasingly rely on code models for drafting implementations, completing functions, explaining APIs, and repairing errors~\cite{peng2023copilot,jain2025livecodebench}.

Despite this progress, code LLMs remain vulnerable to hallucination. 
In natural language, hallucination often appears as fluent but unsupported content~\cite{huang2025hallucination,pearce2022asleep}. This problem is subtle in programming because generated code may look reasonable, compile successfully, and even pass a small number of examples while still failing hidden tests or violating implicit correctness constraints ~\cite{spracklen2025packagehallucination,agarwal2024codemirage,zhang2025practicalhallucination}. 
Prior studies have shown that AI-generated code can contain insecure patterns, package hallucinations, and repository-level inconsistencies, suggesting that correctness failures are not merely surface-level syntax errors but often reflect missing semantic structure~\cite{kim2026sok,ravichander2025halogen,si2026teaching}. 
Therefore, improving code generation requires more than making models fluent in programming syntax; it requires helping them recover the latent structure that makes a program correct.

\iffalse
A growing body of work attempts to mitigate these failures through execution feedback, self-repair, reranking, and iterative refinement~\cite{olausson2024self,wang2026inspectcoder,kamoi2024can,liu2024large}. 
For example, AlphaCode uses large-scale sampling and filtering to select promising candidates~\cite{li2022alphacode}, CodeRL trains code models with execution-based reward signals, Self-Debugging teaches models to inspect and repair their own programs, and Reflexion and Self-Refine use verbal feedback to improve later attempts~\cite{le2022coderl,chen2024selfdebug,shinn2023reflexion,madaan2023selfrefine}. 
These approaches connect generation with feedback from tests, environments, or model-generated critiques. 
However, they still leave a key gap: execution results are often sparse, natural-language critiques can be vague or hallucinate themselves, and passing public tests does not necessarily reveal why a solution is structurally correct~\cite{yan2025guiding,zheng2025processbench,tie2026can}. 
\fi
\begin{figure}
\centering
\includegraphics[width=\linewidth]{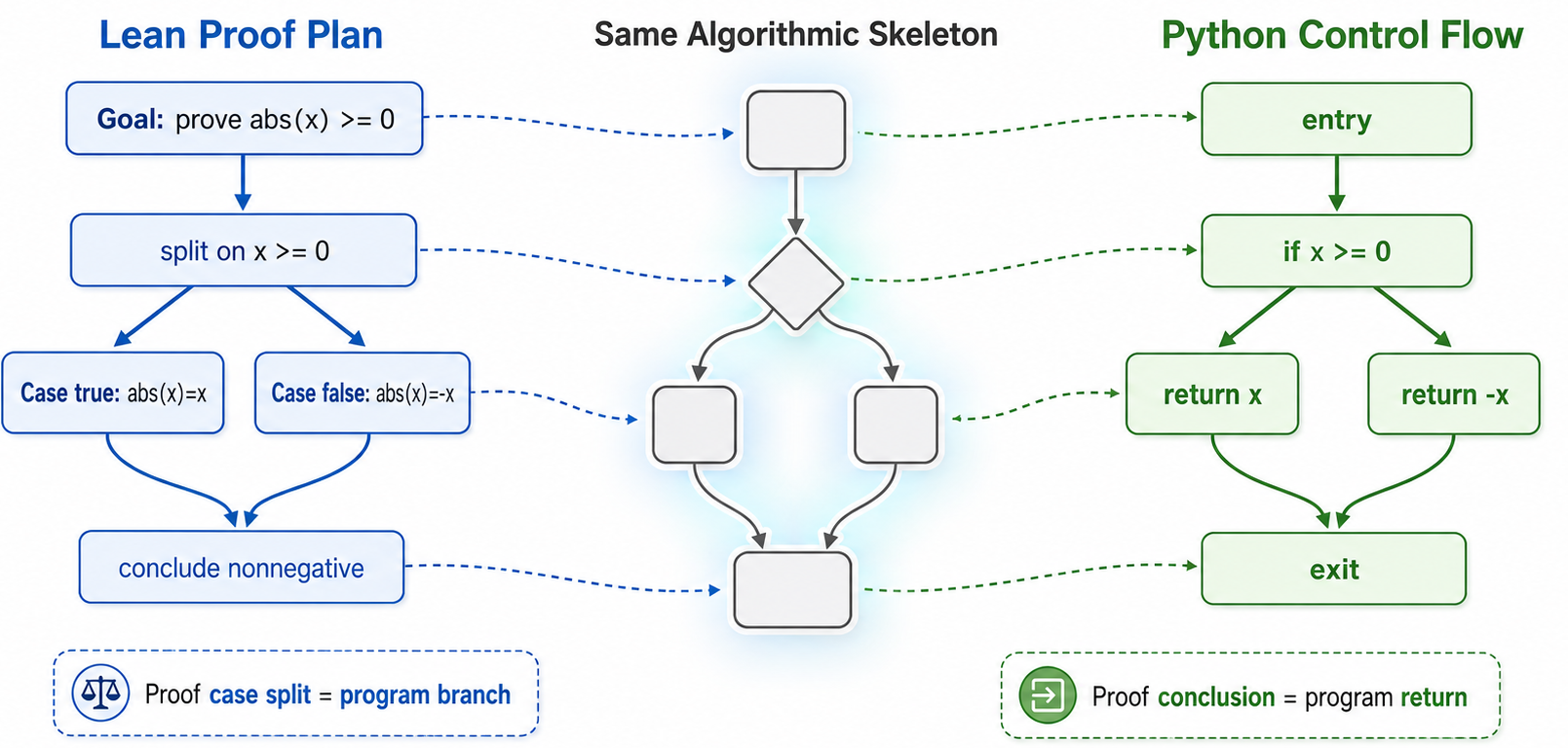}
\caption{\textbf{Shared Algorithmic Skeleton Between Lean Proof and Python Control Flow.} A Lean proof plan exposes case splits, branch obligations, and final conclusions that align with the branch and return structure of an executable Python program.
}
\label{fig:hook}
\end{figure}

Formal proof provides an opportunity to fill this gap because it exposes reasoning as an explicit structure rather than as an implicit byproduct of final code~\cite{demoura2015lean,leino2010dafny}. 
Figure~\ref{fig:hook} illustrates the key observation behind this work: a proof plan for even a simple property, such as nonnegativity of an absolute-value computation, contains a case split and conclusion structure that mirrors the control-flow branches and return sites of the corresponding Python program.
Proof assistants such as Lean decompose a theorem into proof states, subgoals, tactic applications, case splits, obligations, and lemma invocations, which can be viewed as a tactic graph describing how a solution is semantically organized~\cite{demoura2015lean,yang2023leandojo,jiang2023draft,azerbayev2024llemma}. 
Similarly, a Python implementation can be represented as a program graph, such as an abstract syntax tree, control-flow graph, or data-flow graph, where nodes and edges capture how computation proceeds through branches, loops, calls, state updates, and returns~\cite{aho2006compilers,ferrante1987pdg,yamaguchi2014codepropertygraphs}. 
Although Lean tactic graphs and Python control-flow graphs are written in different languages of reasoning, they often share a latent algorithmic structure: proof case splits correspond to program branches, invariant-preservation obligations correspond to loop bodies, inductive decomposition corresponds to recursion, and witness construction corresponds to return-value assembly. 
This structural correspondence suggests that graph similarity can serve as a bridge between formal reasoning and executable program generation, providing process-level supervision that is richer than pass/fail execution feedback but more grounded than free-form natural language critique.

In this paper, we propose \textsc{GraphAlignCoder}, a training framework based on a principle: correct programs have structure, formal obligations make correctness structure explicit, and graph alignment can transfer this structure into code LLM training.
Our approach extracts program graphs from solutions first, including control-flow, data-flow, and region-level structures such as loops and returns. 
In parallel, we construct graphs that encode semantic obligations, tactic skeletons, and subgoals. 
This design gives code LLMs an auditable semantic scaffold for learning the organization of correct programs.
The resulting framework has two parts. First, \toolname{} constructs a filtered structural scaffold by aligning correctness obligations with concrete regions of trusted Python programs. Second, it adapts the model under code generation: the model may learn from structural annotations during training, but the final deployed behavior remains ordinary Python generation from the task alone.

\begin{figure*}
\centering
\includegraphics[width=0.95\textwidth]{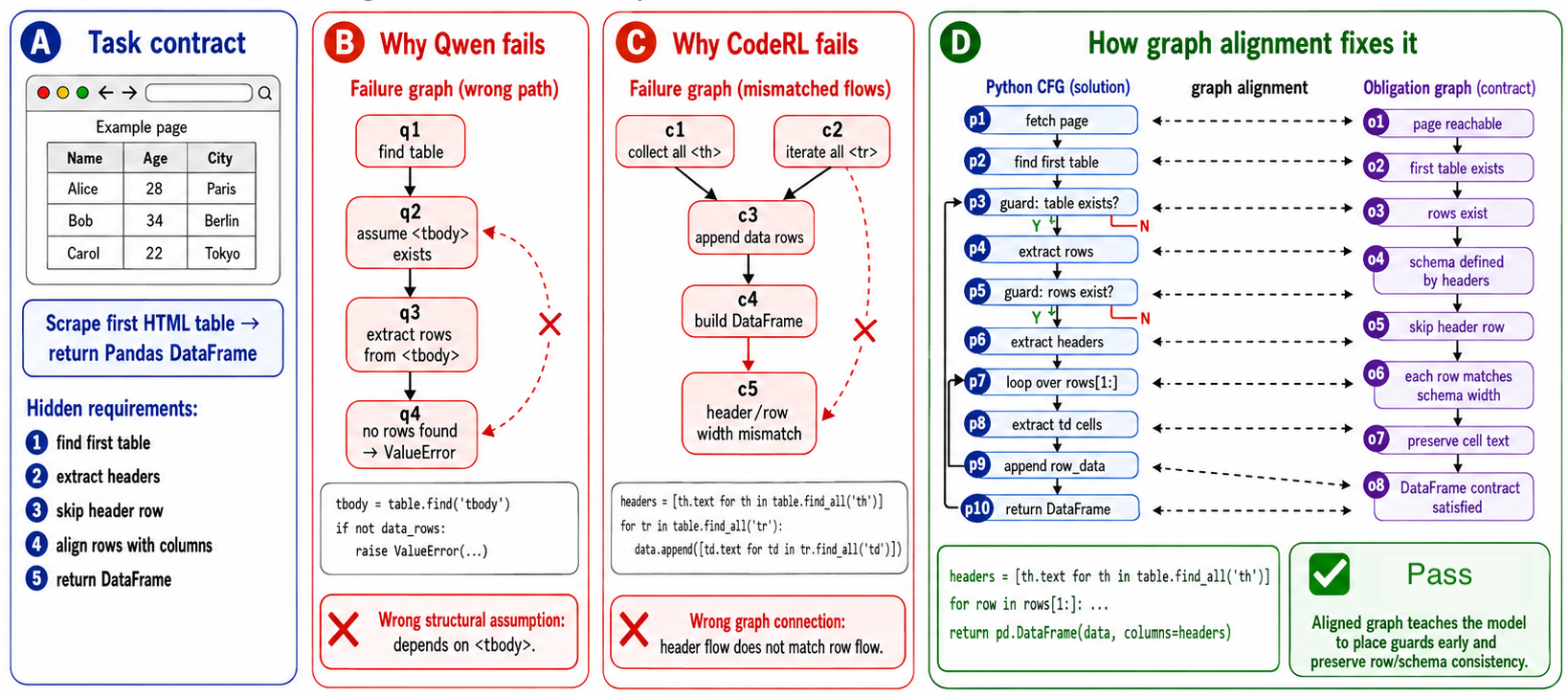}
\caption{\textbf{Motivating Example: Structural Failure and Graph-Aligned Correction.} The task requires scraping the first HTML table into a Pandas DataFrame. The base model fails by assuming a specific \texttt{tbody} structure, and CodeRL fails by mismatching header and row flows. \toolname{} succeeds by aligning the Python control-flow graph with obligation-level constraints such as table existence, row/schema consistency, and DataFrame construction.
}
\label{fig:moti}
\end{figure*}

This paper makes the following contributions:
\begin{itemize}
    \item We introduce \toolname{}, a framework that connects executable program structure with proof-flow structure. It aligns implementation regions and Lean proof transitions, injects this structure into a code model, and consolidates it into Python generation.

    \item \toolname{} consistently outperforms the base model, code-only SFT, and CodeRL across all three benchmarks. Compared with CodeRL, it improves the solved count by 31.6\% on LiveCodeBench v6 (50 versus 38) and 43.8\% on BigCodeBench Hard (23 versus 16).

    \item Ablations validate the complementary roles of structural injection and consolidation. Structural injection increases LiveCodeBench v6 from 33 to 44 solved tasks, while consolidation further raises it to 49 and recovers BigCodeBench Full from 339 to 352.
\end{itemize}

\iffalse
The rest of the paper is organized as follows. 
Section~\ref{sec:motivating-example} presents a motivating example that contrasts structurally invalid and structurally valid generated programs. 
Section~\ref{sec:related-work} reviews code generation, formal proof guidance, and program-graph representations. 
Section~\ref{sec:method} describes the \toolname{} training pipeline. 
Section~\ref{sec:evaluation} defines the benchmarks, baselines, and evaluation protocol, and Section~\ref{sec:results} reports the main results, difficulty/category analysis, and ablation findings.
\fi

\section{Motivating Example}
\label{sec:motivating-example}
\subsection{Motivating Example}

\label{subsec:motivating_example}

Figure~\ref{fig:moti} shows a representative example from BigCodeBench where both the base code model and the CodeRL baseline fail, while \toolname{} succeeds. 
The task asks the model to scrape the first HTML table from a web page and convert it into a Pandas DataFrame. 
Although this task appears to be a simple library-usage problem, it requires the implementation to preserve several implicit obligations: fetch the page safely, locate the first table, extract headers, skip the header row when collecting data, preserve the column--row alignment, and return a DataFrame with the expected schema.

The base model generates plausible code using \texttt{requests}, \texttt{BeautifulSoup}, and \texttt{pandas}, but it assumes that the table body is explicitly wrapped in a \texttt{tbody} element. 
When the benchmark page does not follow this exact structure, the implementation fails with \texttt{ValueError("No table data found.")}. 
The CodeRL baseline repairs part of the structure by iterating over table rows, but it extracts all \texttt{th} elements as headers while also processing rows in a way that creates a mismatch between the number of columns and the number of values. 
As a result, it still fails when constructing the DataFrame.

\toolname{} succeeds because its generated program follows a more obligation-preserving structure. 
It first checks that a table exists, then checks that rows exist, extracts headers from \texttt{th} cells, iterates over data rows after the header row, extracts \texttt{td} cells consistently, and finally constructs the DataFrame using the aligned header and data lists. 
This example illustrates the central motivation of our approach: the difference between failure and success is not merely whether the model knows the right library calls, but whether it organizes those calls according to the semantic structure of the task.

From the Python side, the correct solution contains a control-flow structure with a connection guard, a table-existence guard, a row-existence guard, a header-extraction region, a data-row extraction loop, and a return-construction region. 
From the proof side, the same solution can be described as a set of obligations: the page must be reachable, the target table must exist, headers must define the DataFrame schema, each data row must match the schema, and the returned object must preserve the extracted table content. 
\toolname{} uses this alignment between operational regions and obligations as auxiliary supervision, while still using Python code as the executable target.

\section{Related Work}
\label{sec:related-work}

\subsection{Code Language Models and Code Generation}
\label{sec:rw-code-llms}

Large language models for code have made code generation a central task in software engineering research. Codex showed that pretrained language models can synthesize functional programs from natural-language prompts~\cite{chen2021codex,li2022alphacode}. Recent open code models, including Code Llama and Qwen-Coder, further improve completion, instruction following, and multilingual programming ability through large-scale code pretraining and instruction tuning ~\cite{roziere2023codellama,hui2024qwen25coder}. These advances are increasingly evaluated on benchmarks that stress different forms of programming competence ~\cite{jain2025livecodebench,zhuo2025bigcodebench}.
%Execution-feedback methods connecting generation to behavioral evidence. CodeRL trains code models with execution-based rewards, and self-repair methods such as Self-Debugging, Reflexion, and Self-Refine use debugging traces or verbal feedback to improve later attempts \cite{le2022coderl,chen2024selfdebug,shinn2023reflexion,madaan2023selfrefine}. 

\subsection{Formal Proof Guidance for Reasoning}
\label{sec:rw-formal-proof}

Formal proof assistants expose reasoning as explicit intermediate structure, making them a natural source of supervision for models that must preserve correctness constraints. Lean represents proofs through states, subgoals, tactics, case splits, lemma applications, and obligation discharge, which provides a structured view of how a theorem is solved ~\cite{demoura2015lean}. Recent work has used this structure to train or guide theorem-proving systems: LeanDojo supports retrieval-augmented theorem proving, Draft-Sketch-Prove uses informal proof sketches to guide formal proof search, and Llemma develops language models specialized for mathematical reasoning ~\cite{yang2023leandojo,jiang2023draft,azerbayev2024llemma,ospanov2026apollo}. These systems show that proof artifacts can provide useful process-level signals beyond final answers~\cite{zhang2025codegradintegratingmultistepverification}.
However, theorem-proving supervision does not directly solve Python code generation. Formal proofs operate over explicit logical statements, while programming benchmarks require executable implementations that satisfy tests, APIs, input formats, and hidden edge cases~\cite{cao2025informal,zhou2024don}. 

\subsection{Program Graphs and Structural Code Representations}
\label{sec:rw-program-graphs}

Program graphs provide a long-standing way to represent code beyond token sequences~\cite{allamanis2017learning}. Classical compiler analysis uses abstract syntax, control flow, data flow, and program dependence graphs to capture how computation is organized across branches, loops, definitions, uses, and state updates ~\cite{aho2006compilers,ferrante1987pdg}. More recent work extends these ideas to learning-based code analysis: code property graphs combine syntactic, control-flow, and data-flow information for vulnerability discovery, while GraphCodeBERT incorporates data-flow structure into code representation learning \cite{yamaguchi2014codepropertygraphs,guo2021graphcodebert}. Formal verification provides a complementary structural view: Floyd--Hoare logic attaches assertions to program points and uses branch conditions and loop invariants to derive correctness obligations, while modern verifiers such as Dafny automate these obligations for executable programs~\cite{floyd1967assigning,hoare1969axiomatic,leino2010dafny}.

\section{Methodology}
\label{sec:method}

\begin{figure*}[t]
\centering
\includegraphics[width=0.95\textwidth]{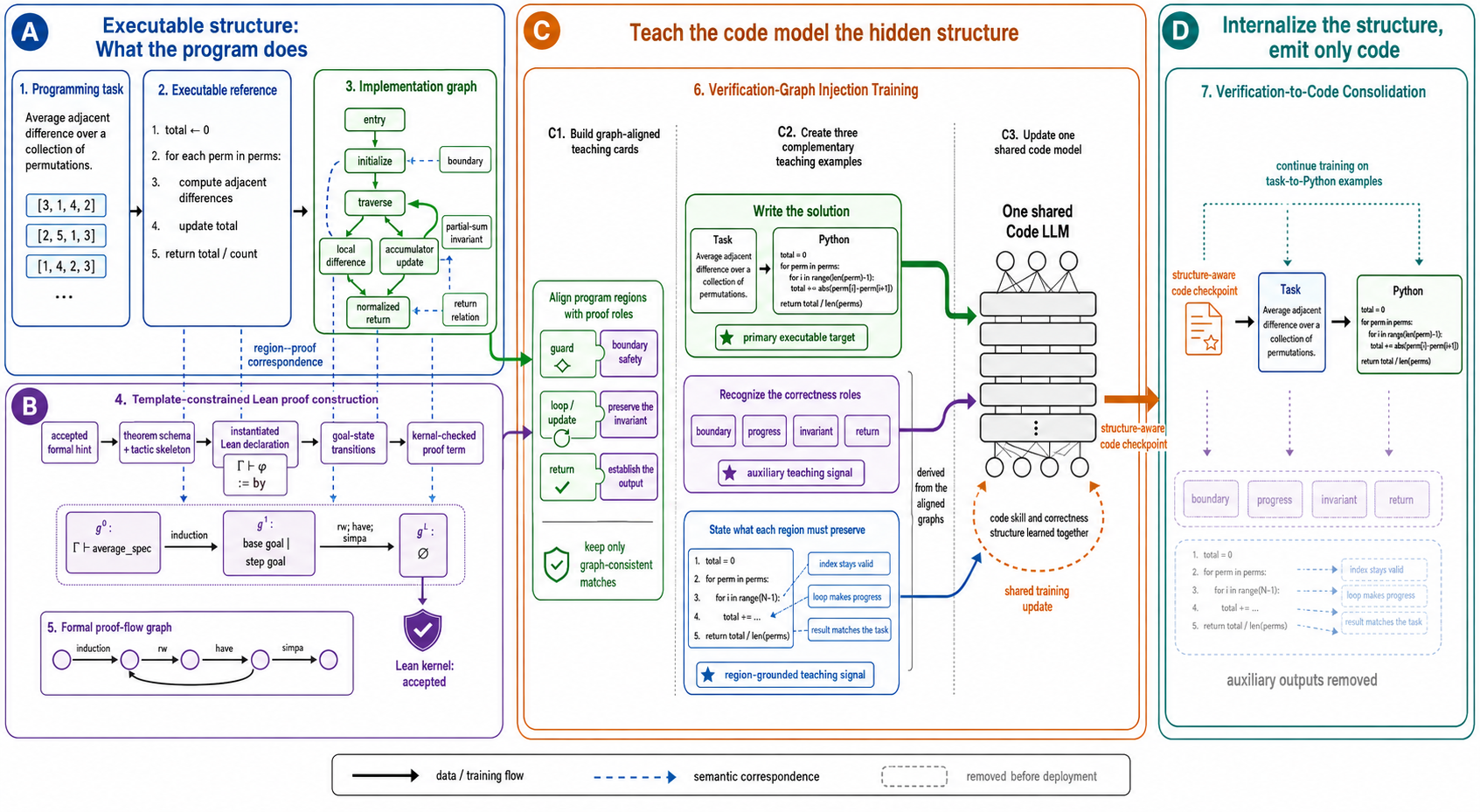}
\caption{\textbf{Overview of \toolname{}.} (A) We parse a solution into an implementation graph whose regions expose boundaries, updates, and return relations. (B) Template-constrained Lean construction turns formal hints into kernel-checked goal transitions and a formal proof-flow graph. (C) Region--proof correspondences produce three complementary training targets---executable code, verification roles, and region-grounded conditions---that update one shared code model. (D) Verification-to-Code Consolidation continues from the structure-aware checkpoint using executable Python alone; auxiliary outputs are removed before deployment.}
\label{fig:framework}
\end{figure*}

\toolname{} follows the four-stage pipeline in Figure~\ref{fig:framework}. Panels A and B expose complementary structures from Python code. Panel C turns its correspondence into Verification-Graph Injection Training. Panel D then removes the auxiliary outputs and consolidates the structure into Python code.

\subsection{Executable Structure: Implementation Graph}
\label{sec:implementation-graph}

For task $i$, let $x_i$ contain the natural-language specification, starter code, and visible tests, and let $y_i$ be a correct Python code snippet provided by the benchmark. Thus, $y_i$ defines what the model should produce, while the graph below exposes how that solution is organized.
We deterministically parse $y_i$ with Python's AST and construct the implementation graph
\begin{equation}
G_i^{I}=\mathcal{G}_{I}(y_i)=(V_i^{I},E_i^{I},R_i),
\end{equation}
where $V_i^{I}$ contains function-entry and statement nodes, $R_i$ groups source spans into regions such as guards, loops, calls, recursive sites, updates, and returns, and $E_i^{I}$ contains control-flow, call, and dominance relations. The graph converts a whole-program target into region-level structure. As visualized in Figure~\ref{fig:framework}A, correctness descriptions can be attached to a specific boundary, update, or return rather than to the program as an undifferentiated sequence.

\subsection{Formal Structure: Lean Proof-Flow Graph}
\label{sec:proof-flow-graph}

A \emph{formal proof-flow graph} contains proof states as nodes and tactic-driven goal transformations as edges. We construct it through two constrained teacher LLM passes with GPT-5.2. The first teacher reads $x_i$, $y_i$, and the region inventory $R_i$ to identify boundary, invariant, progress, and return requirements. The second maps these requirements to a formal hint set
\begin{equation}
H_i^{F}=\operatorname{LLM}_{F}
\left(x_i,y_i,\operatorname{summary}(G_i^{I}),\mathcal{T}\right),
\end{equation}
where the fixed library $\mathcal{T}$ contain theorem templates and tactic skeletons.  Each hint $h_{ij}\in H_i^{F}$ specifies a known theorem shape, tactic plan, proof roles, and lemma families. Restricting the teacher to these libraries makes it select controlled formal building blocks instead of generating an unrestricted Lean program.

Lean turns each accepted hint into the explicit process shown in Figure~\ref{fig:framework}B. First, the generator selects the theorem schema and tactic skeleton whose declared roles match $h_{ij}$:
\begin{align}
\tau_{ij}&=\operatorname{Select}_{\mathcal{T}}(h_{ij}).
\end{align}
Second, it instantiates the schema's types, variables, hypotheses, and conclusion with task-specific symbols. The resulting Lean declaration has context $\Gamma_{ij}$, proposition $\varphi_{ij}$, and a tactic block inserted after \texttt{:= by}:
\begin{equation}
P_{ij}^{\mathrm{Lean}}
=\operatorname{Instantiate}(\tau_{ij},x_i,h_{ij}).
\end{equation}
Then, the tactic engine executes the skeleton as a sequence of state transformations,
\begin{equation}
g_{ij}^{0}\xrightarrow{}g_{ij}^{1}
\xrightarrow{}\cdots
\xrightarrow{}g_{ij}^{L}.
\end{equation}
where each $g_{ij}^{\ell}$ is the current list of unresolved goals. For example, \texttt{cases} creates branch-specific goals, \texttt{induction} creates base and step goals, \texttt{rw} applies a typed equality, \texttt{have} adds a proved fact to $\Gamma_{ij}$, and \texttt{exact} or \texttt{simpa} closes a goal. Successful execution ends with $g_{ij}^{L}=\varnothing$.
 The elaborator compiles a successful trace into a proof term, and Lean's kernel independently checks the term.

Rather than using the proof term as an opaque certificate, we extract the goal-state trajectory that produced it. Let $P_i^{\mathrm{Lean}}=\{P_{ij}^{\mathrm{Lean}}\}_j$ denote the generated artifacts for task $i$; then
\begin{equation}
G_i^{F}=\operatorname{ExtractTacticGraph}
\left(P_i^{\mathrm{Lean}},H_i^{F}\right)=(V_i^{F},E_i^{F}).
\end{equation}
The extractor connects goal states with tactic-labeled edges and records whether transitions perform case analysis, induction, rewriting, invariant propagation, or witness construction. Together, the implementation and proof-flow graphs form the scaffold
\begin{equation}
S_i=(G_i^{I},G_i^{F},H_i^{F})
\end{equation}
that preserves the provenance of both executable regions and formal transitions. The dashed links between Panels A and B represent shared semantic roles.

\subsection{Verification-Graph Injection Training}
\label{sec:contract-adaptation}
\label{sec:structure-sft}

In Panel C, we show three complementary examples and convert these scaffolds into one code model. In step C1, we project the graph nodes and transitions into compact roles. Let $C_i^{I}$ and $C_i^{F}$ denote the roles recovered from the implementation and proof-flow graphs, and let $A_i$ retain their shared semantic support:
\begin{align}
C_i^{I}&=\pi_I\!\left(\operatorname{Roles}(G_i^{I})\right),\\
C_i^{F}&=\pi_F\!\left(\operatorname{Roles}(G_i^{F})\right),\\
A_i&=C_i^{I}\cap C_i^{F}.
\end{align}
We retain only implementation roles supported by the source and use cross-graph overlap as alignment evidence. Controlled schemas then translate each retained role into a region-level condition---for example, a loop update must preserve its invariant, or a return must establish the required output relation:
\begin{align}
C_i&=\operatorname{Filter}_{C}(C_i^{I};A_i,G_i^{I},y_i),\\
\widetilde O_i&=\operatorname{Gen}(C_i,R_i),\\
O_i&=\operatorname{Filter}_{O}(\widetilde O_i;G_i^{I},y_i).
\end{align}
We serialize these accepted sets consistently as
\begin{equation}
c_i=\operatorname{ser}(C_i),
\qquad o_i=\operatorname{ser}(O_i).
\end{equation}

In step C2, each scaffold yields three separate prompt--target pairs: code generation $x_i\rightarrow y_i$, concept prediction $q_i^{C}\rightarrow c_i$, and obligation prediction $q_i^{O}\rightarrow o_i$. The graphs are not appended as privileged text to the Python prompt. Instead, step C3 routes all three examples through the same model parameters. Consequently, predicting an invariant or boundary condition updates the same representation later used by $p_\theta(y_i\mid x_i)$.

For prompt $q$, response $a=(a_1,\ldots,a_T)$, and example weight $w$, prompt and padding tokens are masked and the response-token loss is
\begin{equation}
\ell_\theta(q,a;w)=-\frac{w}{T}\sum_{t=1}^{T}
\log p_\theta(a_t\mid q,a_{<t}).
\end{equation}
The three objectives specialize this common loss to executable tokens, shared verification concepts, and region-grounded obligations:
\begin{align}
\mathcal{L}_{\mathrm{code}}
&=-\sum_i\frac{1}{|y_i|}\log p_\theta(y_i\mid x_i),\\
\mathcal{L}_{\mathrm{concept}}
&=-\sum_i\frac{1}{|c_i|}\log p_\theta(c_i\mid q_i^{C}),\\
\mathcal{L}_{\mathrm{obligation}}
&=-\sum_i\frac{1}{|o_i|}\log p_\theta(o_i\mid q_i^{O}).
\end{align}
The reported injection objective is therefore
\begin{equation}
\mathcal{L}_{\mathrm{inject}}
=\mathcal{L}_{\mathrm{code}}
+\mathcal{L}_{\mathrm{concept}}
+\mathcal{L}_{\mathrm{obligation}}.
\end{equation}
%Concept and obligation gradients organize the shared representation without replacing executable code as the primary learning signal. This is loss-level graph injection rather than a graph neural network or graph-conditioned inference.

\subsection{Verification-to-Code Consolidation}
\label{sec:verification-consolidation}

Finally, we convert the structure-aware model back to the output contract required at deployment. After injection, we stop requesting concept and obligation responses and continue training the same checkpoint on task-to-Python examples. This removes the auxiliary output channels while carrying their parameter updates into the code-only phase.

Let $\theta_0$ denote the initial trainable parameters and $\theta_1$ the checkpoint produced by Verification-Graph Injection Training. For the consolidation set $\mathcal{D}_{\mathrm{con}}=\{(x_i,y_i)\}_{i=1}^{N_{\mathrm{con}}}$, where $y_i=(y_{i,1},\ldots,y_{i,T_i})$ is an executable reference implementation, we initialize $\theta\leftarrow\theta_1$ and optimize the code loss
\begin{equation}
\mathcal{L}_{\mathrm{con}}(\theta)
=-\frac{1}{N_{\mathrm{con}}}
\sum_{i=1}^{N_{\mathrm{con}}}\frac{1}{T_i}
\sum_{t=1}^{T_i}
\log p_{\theta}
\!\left(y_{i,t}\mid x_i,y_{i,<t}\right).
\label{eq:consolidation-loss}
\end{equation}

where $T_i$ is the target length and $y_{i,<t}$ is the target prefix. Optimization yields $\theta_2$, the consolidated checkpoint shown at the end of Panel D.

Let $S_{\mathrm{I}}$ and $S_{\mathrm{C}}$ be the numbers of injection and consolidation steps. The two-phase curriculum can be written as
\begin{equation}
\mathcal{L}_{\mathrm{curr}}^{(s)}(\theta)=
\begin{cases}
\mathcal{L}_{\mathrm{inject}}(\theta),
&1\leq s\leq S_{\mathrm{I}},\\[1mm]
\mathcal{L}_{\mathrm{con}}(\theta),
&S_{\mathrm{I}}<s\leq S_{\mathrm{I}}+S_{\mathrm{C}}.
\end{cases}
\label{eq:two-phase-curriculum}
\end{equation}

Injection and consolidation therefore serve complementary purposes. Injection makes boundaries, invariants, progress conditions, and return relations explicit; consolidation rewards the model for realizing the same organization in Python. At inference, $\theta_2$ receives the ordinary programming input and emits one Python solution, with no implementation graph or proof-flow graph.

\section{Evaluation}
\label{sec:evaluation}

\subsection{Benchmarks and Inference Protocol}
\label{sec:benchmarks-metrics}
We evaluate on two complementary code-generation benchmarks. LiveCodeBench v6 contains 175 held-out algorithmic programming tasks and tests generalization beyond the LiveCodeBench v1--v4 tasks used for training. BigCodeBench Complete contains 1,140 function-level tasks emphasizing API usage, dependency handling, data manipulation, and realistic function completion; its Hard subset contains 148 tasks with stricter functional requirements. BigCodeBench is used only for evaluation. The main metric is Pass@1 under the single-sample greedy contract above. All compared models follow the same evaluation contract: each receives the ordinary programming-task input and produces one Python solution with greedy decoding. Pass@1 is the fraction of tasks for which the single generated program passes the official execution suite.

\subsection{Baselines and Ablations}
\label{sec:baselines}
\subsubsection{Baselines Approach}
We compare \textsc{GraphAlignCoder} against base-model and execution-feedback baselines. \textbf{Base Model} is the original Qwen code model. This baseline measures how much performance is already present in the pretrained code model. \textbf{Code-only SFT (Supervised Fine-Tuning)} is a diagnostic checkpoint trained only on Python implementations. 
\textbf{CodeRL} denotes the reported execution-feedback baseline, which improves code generation using execution outcomes. It is evaluated with the same input, decoding, and one-sample contract as the other models~\cite{le2022coderl}.

\subsubsection{Ablation Variants}
\noindent\textbf{Verification-graph injection} is a diagnostic checkpoint that adds the active auxiliary concept-prediction and obligation targets to the primary executable-code target. This measures loss-level graph injection.

%\noindent\textbf{Consolidation-Only} is a diagnostic checkpoint that removes all auxiliary targets after injection and continues on executable Python outputs. This row tests whether the learned verification structure can be consolidated into the code-only output distribution.

\noindent\textbf{Consolidation-only} is initialized directly from the base model and trained to generate executable Python from tasks augmented with graph-derived verification information, such as region-level obligations. 

%Unlike verification-graph injection, it does not predict concepts or obligations as auxiliary outputs; unlike code-only SFT, it still receives structural guidance during training, thereby testing whether the ``why'' of correctness can be translated directly into Python code.

%\textbf{\textsc{GraphAlignCoder}} is the proposed model trained with executable reference implementations, implementation-graph-derived verification concepts, and region-grounded verification obligations. Its formal proof-flow artifacts support scaffold construction but do not receive a separate nonzero loss in the reported configuration. At test time, it receives exactly the same ordinary task input as the baselines.

\subsection{Category and Difficulty Annotation}
\label{sec:category-annotation}

We use a separate GPT-5.2 annotation pipeline only for analysis grouping. The script batches evaluation tasks, sends compact task summaries to the GPT-5.2 endpoint with temperature $0$, and requires output containing the task identifier, one allowed category, an estimated difficulty label, a confidence score, and a short rationale. For LiveCodeBench, the allowed categories are algorithmic families such as array, string, dynamic programming and others. For BigCodeBench, the allowed categories are function-oriented groups such as data science, plotting/visualization, filesystem/OS, web/text processing, crypto/encoding and other.

\subsection{Training and Hardware Setting}
\label{sec:training-setting}
All \toolname{} checkpoints are initialized from \texttt{Qwen/Qwen3.5-4B-Base} and trained with LoRA adapters. The training data follows the executable-reference policy defined in Methodology. Verification-to-Code Consolidation contains 514 code-only examples and runs for 65 optimization steps.  All checkpoints use LoRA rank 16, and the final checkpoint uses adapter tensor scale $0.55$. Training and checkpoint construction run on a local workstation with $2\times$ NVIDIA RTX A6000 GPUs. Final results are computed from saved greedy generations with the official LiveCodeBench and BigCodeBench evaluators.

\begin{table}[t]
\centering
\caption{Overall Pass@1 performance. Values are solved counts with rates in parentheses; $N$ is the benchmark size.}
\label{tab:overall_performance}
{\scriptsize
\setlength{\tabcolsep}{1.0pt}
\renewcommand{\arraystretch}{0.88}
\begin{tabular*}{\linewidth}{@{\extracolsep{\fill}}lrcccc@{}}
\toprule
\textbf{Bench.} & \textbf{N} & \textbf{Base} & \textbf{Code-only SFT} & \textbf{CodeRL} & \textbf{GAC} \\
\midrule
LCB v6
& 175
& \score{15}{.086}
& \score{33}{.189}
& \score{38}{.217}
& \best{\score{50}{.286}} \\

BCB Hard
& 148
& \score{9}{.061}
& \score{15}{.101}
& \score{16}{.108}
& \best{\score{23}{.155}} \\

BCB Full
& 1140
& \score{209}{.183}
& \score{354}{.311}
& \score{359}{.315}
& \best{\score{363}{.318}} \\
\bottomrule
\end{tabular*}
}
\end{table}

\begin{table}[t]
\centering
\caption{Difficulty-level Pass@1 by bucket. Values are solved count with rate in parentheses}
\label{tab:difficulty_analysis}
{\scriptsize
\setlength{\tabcolsep}{1.4pt}
\renewcommand{\arraystretch}{0.88}
\begin{tabular*}{\linewidth}{@{\extracolsep{\fill}}lrccc@{}}
\toprule
\textbf{Diff.} & \textbf{N} & \textbf{Code-only SFT} & \textbf{CodeRL} & \textbf{GAC} \\
\midrule
\multicolumn{5}{@{}l}{\textit{LiveCodeBench v6}} \\
Easy
& 43
& \score{25}{.581}
& \score{25}{.581}
& \best{\score{32}{.744}} \\
Medium
& 52
& \score{6}{.115}
& \score{9}{.173}
& \best{\score{10}{.192}} \\
Hard
& 80
& \score{2}{.025}
& \score{4}{.050}
& \best{\score{8}{.100}} \\
\midrule
\multicolumn{5}{@{}l}{\textit{BigCodeBench}} \\
Easy
& 396
& \best{\score{164}{.414}}
& \score{163}{.412}
& \score{160}{.404} \\
Medium
& 638
& \score{178}{.279}
& \score{185}{.290}
& \best{\score{189}{.296}} \\
Hard
& 106
& \score{12}{.113}
& \score{11}{.104}
& \best{\score{14}{.132}} \\
\bottomrule
\end{tabular*}
}
\end{table}

\begin{table}[t]
\centering
\caption{Representative category-level Pass@1. }
\label{tab:category_analysis}
{\scriptsize
\setlength{\tabcolsep}{1.0pt}
\renewcommand{\arraystretch}{0.83}
\begin{tabular*}{\linewidth}{@{\extracolsep{\fill}}lrccc@{}}
\toprule
\textbf{Category} & \textbf{N} & \textbf{Code-only SFT} & \textbf{CodeRL} & \textbf{GAC} \\
\midrule
\multicolumn{5}{@{}l}{\textit{LiveCodeBench v6}} \\
Array
& 10
& \score{5}{.500}
& \score{4}{.400}
& \best{\score{7}{.700}} \\
DP
& 35
& \score{0}{.000}
& \score{0}{.000}
& \best{\score{1}{.029}} \\
Graph
& 19
& \score{2}{.105}
& \score{2}{.105}
& \best{\score{4}{.211}} \\
Simulation
& 12
& \score{3}{.250}
& \best{\score{4}{.333}}
& \best{\score{4}{.333}} \\
Sorting
& 5
& \best{\score{2}{.400}}
& \best{\score{2}{.400}}
& \score{1}{.200} \\
String
& 19
& \score{9}{.474}
& \score{10}{.526}
& \best{\score{12}{.632}} \\
\midrule
\multicolumn{5}{@{}l}{\textit{BigCodeBench}} \\
API/Lib
& 84
& \best{\score{34}{.405}}
& \score{33}{.393}
& \score{30}{.357} \\
Crypto
& 47
& \score{12}{.255}
& \score{12}{.255}
& \best{\score{15}{.319}} \\
FS/OS
& 186
& \score{45}{.242}
& \score{49}{.263}
& \best{\score{57}{.306}} \\
Plot/Vis.
& 264
& \score{79}{.299}
& \score{80}{.303}
& \best{\score{81}{.307}} \\
Web/Text
& 88
& \score{19}{.216}
& \score{19}{.216}
& \best{\score{26}{.295}} \\
\bottomrule
\end{tabular*}
}
\end{table}

\begin{table}[t]
\centering
\caption{Deployed code-generation performance. }
\label{tab:rq3_deployed}
{\scriptsize
\setlength{\tabcolsep}{1.0pt}
\renewcommand{\arraystretch}{0.88}
\begin{tabular*}{\linewidth}{@{\extracolsep{\fill}}lccc@{}}
\toprule
\textbf{Model}
& \textbf{LCB v6}
& \textbf{BCB}
& \textbf{BCB-H} \\
\midrule
Code-only SFT
& \score{33}{.189}
& \score{354}{.311}
& \score{15}{.101} \\

\toolname{}-Structural injection
& \score{44}{.251}
& \score{339}{.297}
& \score{17}{.115} \\

\toolname{}-Consolidation-Only
& \score{49}{.280}
& \score{352}{.309}
& \score{19}{.128} \\

\toolname{}
& \best{\score{50}{.286}}
& \best{\score{363}{.318}}
& \best{\score{23}{.155}} \\
\bottomrule
\end{tabular*}
}
\end{table}

\section{Results}
\label{sec:results}

\subsection{RQ1: How does \toolname{} compare with the base model, code-only SFT, and CodeRL?}
\label{sec:rq1}

Table~\ref{tab:overall_performance} demonstrates that \toolname{} consistently achieves the best Pass@1 across all three evaluation settings. On LiveCodeBench v6, \toolname{} solves 50 tasks, compared with 15 for the base-matched model, 33 for code-only SFT, and 38 for CodeRL. Thus, \toolname{} solves more than three times as many tasks as the base model and improves the solved count over code-only SFT and CodeRL by 51.5\% and 31.6\%, respectively. The improvement over both the base and adapted baselines shows that the gain cannot be explained solely by the pretrained model's existing capability or by additional code-only training.

The improvement is similarly pronounced on BigCodeBench Hard. \toolname{} solves 23 tasks, compared with 9 for the base model, 15 for code-only SFT, and 16 for CodeRL. This corresponds to 155.6\% more solved tasks than the base model, 53.3\% more than code-only SFT, and 43.8\% more than CodeRL. 
On BigCodeBench Full, \toolname{} solves 363 tasks, exceeding the base model's 209 solved tasks and improving over code-only SFT and CodeRL, which solve 354 and 359 tasks, respectively. \toolname{} retains their general function-level capability while achieving much larger gains on LiveCodeBench v6 and BigCodeBench Hard. 

\subsection{RQ2: Where Does \toolname{} Generalize? Analysis Across Difficulty Levels and Task Categories}
\label{sec:rq2}

Table~\ref{tab:difficulty_analysis} shows that the LiveCodeBench improvement extends across all difficulty levels. On easy problems, \toolname{} solves 32 of 43 tasks, seven more than either baseline. On medium problems, it achieves the best result with 10 solved tasks. Most notably, \toolname{} solves 8 hard tasks, twice as many as CodeRL and four times as many as code-only SFT. 
A similar advantage appears in the more demanding portions of BigCodeBench. Although \toolname{} is slightly below the baselines on easy tasks, it obtains the best results on both medium and hard tasks, solving 189 and 14 problems, respectively. In particular, it improves over CodeRL from 11 to 14 solved tasks in the hard bucket, corresponding to a 27.3\% relative gain. Thus, the overall improvement of \toolname{} is  concentrated in the settings where maintaining structural consistency is more difficult. 

The category-level results in Table~\ref{tab:category_analysis} provide further evidence for this interpretation. On LiveCodeBench v6, \toolname{} achieves clear improvements in array, graph, and string problems. On BigCodeBench, the largest gains over CodeRL occur in filesystem/OS and web/text processing, where \toolname{} solves 8 and 7 additional tasks, respectively. It also improves in crypto/encoding and plotting/visualization. These categories frequently require a sequence of dependent operations---such as preserving intermediate state and constructing a valid return object---that closely matches the region-level obligations used by our method.
Its strongest gains consistently appear in categories where correct execution depends on organizing operations into a coherent, constraint-preserving program. 

\subsection{RQ3: Which Training Components Drive the Performance Gains of \toolname{}?}
\label{sec:rq3}

\paragraph{Effect of structural injection.}
Adding graph-aligned structural supervision to code-only SFT substantially improves LiveCodeBench v6, increasing the solved count from 33 to 44 and Pass@1 from 18.9\% to 25.1\%. It also improves BigCodeBench Hard from 15 to 17 solved tasks. These gains demonstrate that verification concepts and region-grounded obligations provide useful supervision beyond executable-code imitation. In particular, the 11-task improvement on LiveCodeBench indicates that structural injection helps the model organize algorithmic solutions more effectively.
Structural injection alone, however, reduces BigCodeBench Full performance from 354 to 339 solved tasks. This result highlights an important role of the subsequent training stages: directly introducing auxiliary structural targets improves structure-sensitive generation, but can temporarily shift the model away from its original Python-only output distribution. Thus, structural supervision is effective, but it must be consolidated before the resulting checkpoint can serve as a broadly capable code generator.

\paragraph{Effect of Verification-to-Code Consolidation.}
Consolidation-only tests whether graph-derived obligations can directly guide Python generation without verification-graph injection. Compared with code-only SFT, which learns only from task-to-code pairs, consolidation-only improves LiveCodeBench v6 from 33 to 49 solved tasks and BigCodeBench Hard from 15 to 19, while remaining nearly unchanged on BigCodeBench Full (352 versus 354). These results indicate that exposing the base model to explicit correctness obligations, even when Python is the only training output, improves algorithmic and hard-task generation.
Consolidation-only is nevertheless weaker than the complete \toolname{} model, which reaches 50, 363, and 23 solved tasks on LiveCodeBench v6, BigCodeBench Full, and BigCodeBench Hard, respectively. The full model exceeds consolidation-only by 1, 11, and 4 tasks, showing that obligation-guided Python generation is beneficial but cannot replace the richer representation learned through verification-graph injection. The best performance arises when auxiliary structural learning and consolidation are combined.

\section{Conclusion}
\label{sec:conclusion}

This paper presented \toolname{}, a training framework that uses graph-aligned correctness obligations to improve code generation. The key idea is to connect the executable structure of Python programs with the correctness structure. The model first learns executable code together with graph-derived descriptions of why program regions are correct, and then consolidates this knowledge into the Python-only output distribution required at deployment.

\toolname{} solves 50/175 LiveCodeBench v6 tasks, 23/148 BigCodeBench Hard tasks, and 363/1140 BigCodeBench Full tasks, outperforming both code-only SFT and CodeRL. The ablation study shows that structural injection improves LiveCodeBench v6 from 33 to 44 solved tasks, while consolidation-only provides further benefits by reaching 49 LiveCodeBench and 19 BigCodeBench Hard tasks. However, it remains below the complete \toolname{} model across the three settings.
\toolname{} offers a practical way to transfer mechanically checked correctness structure into ordinary code generation. Future work can extend this framework to richer obligations, additional programming languages, and repository-level generation.

\bibliography{aaai2027}


\end{document}